\documentclass[12pt,psfig]{article}
\usepackage{amssymb}
\usepackage{amsmath,amsthm}
\usepackage{amsfonts}
\usepackage{graphicx}
\usepackage{graphics}
\numberwithin{equation}{section}

\usepackage{hyperref}

\begin{document}
\begin{center}\Large\textbf{Scattering of the Kalb-Ramond
State from a Dynamical D$p$-brane with Background Fields}
\end{center}
\vspace{0.75cm}
\begin{center}{\large Davoud Kamani
and Elham Maghsoodi}\end{center}
\begin{center}
\textsl{\small{Physics Department, Amirkabir University of
Technology (Tehran Polytechnic)\\
P.O.Box: 15875-4413, Tehran, Iran\\
e-mails: kamani@aut.ac.ir , el.maghsoodi@aut.ac.ir \\}}
\end{center}
\vspace{0.5cm}

\begin{abstract}

We apply the boundary state method and operator formalism to 
obtain tree-level scattering amplitude of the 
Kalb-Ramond state from a D$p$-brane. The brane has
a tangential dynamics, and it has been dressed by 
the antisymmetric tensor field, a $U(1)$ internal 
gauge potential and an open string tachyon field.
By using the scattering amplitudes we acquire 
two DBI-like actions corresponding to the target branes. 
Our calculations are in the framework of the 
bosonic string theory. 

\end{abstract}



\newpage

\section{Introduction}

There are various properties of the strings and D-branes, 
such as the non-perturbative string theory,
the D-brane thickness and extension of 
the D-brane effective action, that can be revealed 
via the following scattering 
processes: brane-brane, string-string and string-brane. 
In fact, the brane-brane scattering 
is drastically complicated, because it is
entirely non-perturbative phenomenon \cite{1}.
However, development of 
string theory at a non-perturbative level has shown
that scattering amplitude not only is
related to the string-string collisions \cite{2},
but also it comprises the 
scattering of closed strings from D-branes 
\cite{3}-\cite{15}.
This demonstrates that the string-brane scattering
provides a remarkable insight on 
the non-perturbative string theory.  
Besides, one of the main tools to extract some 
terms of the D-brane effective action is
scattering amplitude of a string from that D-brane
\cite{5}-\cite{9}. 
In addition, the string-brane 
scattering conveniently gives the
key for understanding some essential phenomena 
such as the elucidation of the sizes of the branes and strings.

On the other hand we have boundary states
which represent the D-branes.
A boundary state, which is a closed string state,
accurately encodes all physical properties 
of the corresponding D-brane. This state elaborates that a 
D-brane appears as a source (sink) for emission 
(absorption) of any closed 
string state. Thus, this adequate formalism  
has been widely applied for
various setups of the D-branes \cite{16}-\cite{33},
and scattering of strings from the D-branes \cite{1}-\cite{15}.

In this paper we shall investigate the elastic 
scattering of a specific massless closed string state,
i.e. the Kalb-Ramond state, from a single 
D$p$-brane. The brane has tangential 
rotation and linear motion, 
and has been furnished with a $U(1)$ gauge potential,
an antisymmetric tensor field and an
open string tachyon field. For this purpose
we shall apply the string operator formalism
and boundary state method in the framework  
of the bosonic string theory. The 
scattering amplitudes enable us to extract two DBI-like
actions which are corresponding to the stable 
and unstable dynamical target branes.

In fact, each background field
and the brane dynamics effectively impose a potential 
to the incoming and outgoing string states,
and hence the scattering process is completely 
influenced by them. Therefore, the  
background fields and the brane dynamics   
motivated and stimulated us to compute scattering of a closed 
string state from a dressed-dynamical D$p$-branes.
In other words, simultaneous application of the 
background fields and tangential dynamics prominently
enriches the parametric structure of the scattering process 
and enables us to adjust the strength of the amplitude.
Note that at least one of the background fields should be 
introduced for breaking the Lorentz symmetry inside the 
brane worldvolume to receive a meaningful tangential 
dynamics. Besides, comparison of the scatterings 
from stable and unstable dynamical branes motivated us to 
introduce a tachyonic background too. In addition,
for preserving the conformal symmetry the Kalb-Ramond 
state has been chosen as incident and scattered states.
Furthermore, since this state is massless
and has antisymmetric polarization tensor it
simplifies the scattering equations. 
Finally, obtaining effective actions for the 
dressed-dynamical stable and unstable target branes
completes our motivation for choosing the 
foregoing setup. 

This paper is organized as follows. In Sec. 2,
we introduce a boundary state which is 
associated to a non-stationary D$p$-brane 
with background fields. 
In Sec. 3, we give a brief review of the operator
formalism for scattering of closed strings from a D$p$-brane.  
Then, we calculate the 
scattering amplitude of the Kalb-Ramond state 
from our D$p$-brane. In Sec. 4, we obtain two
DBI-like actions via the scattering amplitudes.
Section 5 is devoted to the conclusions.

\section{Boundary state of a dynamical-dressed D$p$-brane}

For computing the boundary state corresponding 
to a rotating-moving D$p$-brane with 
the internal and background fields 
we begin with the following action for closed string
\begin{eqnarray}
S=&-&\frac{1}{4\pi\alpha'}\int_{\Sigma} d^2\sigma
\left(\sqrt{-g} g^{ab}G_{\mu\nu} \partial_{a}X^\mu
\partial_b X^{\nu} +\epsilon^{ab}B_{\mu\nu}\partial_a X^{\mu}
\partial_b X^{\nu}\right)
\nonumber\\
&+&\frac{1}{2\pi\alpha'}
\int_{\partial\Sigma}d\sigma \left( A_{\alpha}
\partial_{\sigma}X^{\alpha}
+\omega_{\alpha\beta}J^{\alpha\beta}_{\tau}
+T^2 ( X^{\alpha}) \right)~,
\end{eqnarray}
where the set 
$\{x^{\alpha}|\alpha=0,1,\ldots,p\}$ specifies the 
directions along the D$p$-brane 
worldvolume. $\Sigma$ and $\partial\Sigma$ indicate the 
worldsheet of the closed string 
and its boundary, respectively. 
$g_{ab}$ and $G_{\mu\nu}$ are the metrics
of the worldsheet and spacetime. We consider
flat spacetime with the metric 
$G_{\mu\nu}=\eta_{\mu\nu}
={\rm diag}(-1,1,\ldots,1 )$.
For the $U(1)$ gauge field, which lives in the brane
worldvolume, we utilize the gauge $A_{\alpha}
=-\frac{1}{2}F_{\alpha \beta}X^{\beta}$
with the constant field strength $F_{\alpha \beta}$. 
We apply a constant Kalb-Ramond field $B_{\mu\nu}$, and 
the tachyon profile is chosen as
$T^2(X) = \frac{1}{2}U_{\alpha \beta}
X^{\alpha}X^{\beta}$ where the matrix 
$U_{\alpha \beta}$ is symmetric and constant \cite{34}, \cite{35}. 
The tangential linear motion and rotation of 
the brane are given by the antisymmetric 
constant angular velocity $\omega_{\alpha \beta}$,  
and the angular momentum density is
$J^{\alpha \beta}_\tau = X^\alpha \partial_\tau X^\beta
-X^\beta \partial_\tau X^\alpha$.
Note that because of the internal and background fields 
the Lorentz symmetry in the 
worldvolume of the brane has been explicitly lost.
Hence, this tangential dynamics is meaningful.

Vanishing the variation of the action defines the 
equation of motion 
and the following equations for the boundary state
\begin{eqnarray}
&~&\left(\mathcal{K}_{\alpha\beta}
\partial_{\tau}X^{\beta}+\mathcal{F}_{\alpha\beta}
\partial_\sigma X^{\beta}
+ B_{\alpha i}\partial_\sigma X^i
+ U_{\alpha \beta }
X^{\beta}\right)_{\tau=0}|B_x\rangle=0~,
\nonumber\\
&~&\left(X^i-y^i\right)_{\tau=0}|B_x\rangle=0~,
\end{eqnarray}
where $\mathcal{K}_{\alpha
\beta}=\eta_{\alpha\beta}+4\omega_{\alpha\beta}$,
and 
$\mathcal{F}_{\alpha\beta}=B_{\alpha\beta}-F_{\alpha\beta}$
is the total field strength.
The set $\{x^i| i = p+1, \ldots, 25\}$ represents 
the vertical directions to the 
brane worldvolume, and the parameters 
$\{y^i|i = p+1, \ldots,25\}$ exhibit the brane location.

Introducing the solution of the equation of motion 
$(\partial^2_\tau - \partial^2_\sigma)X^\mu (\sigma , \tau)=0$,
i.e, 
\begin{eqnarray}
X^\mu (\sigma , \tau)=x^\mu+2\alpha'p^\mu
\tau+\frac{i}{2}\sqrt{2\alpha'}
\sum_{m\neq 0}\frac{1}{m}
\left(\alpha_m^{\mu}
e^{-2im(\tau-\sigma)}+\tilde{\alpha}_m^\mu
e^{-2im(\tau+\sigma)}\right),
\nonumber
\end{eqnarray}
into Eqs. (2.2) yields these
equations in terms of the closed string 
oscillators and zero modes. Solutions of 
the resulted equations are denoted by the product  
$|B_x\rangle=|B\rangle^{(0)} \otimes|B\rangle^{\rm (osc)}$
where the partial states have the features \cite{33},
\begin{eqnarray}
|B\rangle^{(0)}&=&\frac{T_p}{2\sqrt{\det(U/2)}}
\int_{-\infty}^{\infty}
\exp\bigg{[}i\alpha' \sum_{\alpha \neq \beta}
\left(U^{-1}\mathcal{K}
+\mathcal{K}^T U^{-1}\right)_{\alpha\beta}\;
p^{\alpha}p^{\beta}
\nonumber\\
&+& \frac{i}{2}\alpha' \left(U^{-1}\mathcal{K}
+\mathcal{K}^T U^{-1}\right)_{\alpha \alpha}\;
(p^{\alpha})^2\bigg{]}
\prod^{p}_{\alpha =0}
\left(|p^{\alpha}\rangle dp^{\alpha}\right)
\nonumber\\
&\times& \prod^{25}_{i=p+1} 
\left[\delta \left({x}^i-y^i\right)
|p^i=0 \rangle \right]~,
\end{eqnarray}
\begin{eqnarray}
|B\rangle^{\rm (osc)}
=\prod_{n=1}^{\infty}[\det{M_{(n)}}]^{-1}
\exp\left[{-\sum_{m=1}^{\infty}
\left(\frac{1}{m}\alpha_{-m}^{\mu}S_{(m)\mu\nu}
\tilde{\alpha}_{-m}^{\nu}\right)}\right]
|0\rangle_\alpha|0\rangle_{\tilde{\alpha}}~,
\label{aos}
\end{eqnarray}
where $T_p$ is the tension of the D$p$-brane.
The momentum and center-of-mass position 
of the closed string, in the worldvolume subspace, 
possess the significant relation
\begin{eqnarray}
p^{ \alpha} =-\frac{1}{2\alpha'} 
\left(\mathcal{K}^{-1}U\right)^{ \alpha}_{\;\;{\beta}} 
x^{\beta}.
\nonumber
\end{eqnarray}
The matrix $S_{(m)}$ is defined by
\begin{eqnarray}
S_{(m)\mu\nu}&=&\left((M_{(m)}^{-1}
N_{(m)})_{\alpha\beta},-\delta_{ij}\right)~,
\nonumber\\
M_{(m)\alpha\beta}&=&\mathcal{K}
_{\alpha\beta}- \mathcal{F}_{\alpha\beta}
+\dfrac{i}{2m}U_{\alpha\beta}~,
\nonumber\\
N_{(m)\alpha\beta}&=&\mathcal{K}_{\alpha\beta}
+ \mathcal{F}_{\alpha\beta}-\dfrac{i}{2m}U_{\alpha\beta}~.
\end{eqnarray}
In fact, receiving the solution (2.4) through the 
coherent state method imposes the following conditions  
on the input parameters
$\{B_{\alpha\beta}, F_{\alpha\beta}, U_{\alpha\beta},
\omega_{\alpha\beta}\}$,
\begin{eqnarray}
&~& \eta \mathcal{F} - \mathcal{F} \eta 
+4(\omega \mathcal{F} +\mathcal{F}\omega)=0,
\nonumber\\
&~& \eta U -U\eta +4(\omega U+ U\omega ) =0.
\end{eqnarray}

Since we shall use the covariant formulation
the conformal ghosts should be also introduced.
Thus, we apply the total boundary state
$|B \rangle = |B\rangle^{(0)} \otimes|B\rangle^{\rm (osc)}
\otimes |B \rangle^{\rm (gh)}$, where the ghost
part is the following state \cite{36},  
\begin{equation}
|B\rangle^{\rm (gh)}=\exp{\left[\sum_{m=1}^{\infty}
(c_{-m}\tilde{b}_{-m}
-b_{-m} \tilde{c}_{-m})\right]}\frac{c_0+\tilde{c}_0}{2}
|q=1\rangle|\tilde{q}=1\rangle~.
\end{equation}
This state is independent of the background fields
and the brane dynamics.
\section{Scattering amplitude}

\subsection{Scattering of many strings from a D$p$-branes}

Each brane manifestly couples to all closed string 
states through its corresponding boundary state.
This implies that a D-brane is a source (sink) for 
emitting (absorbing) any closed string state.
The source-sink property of the 
D-branes gives an essential role to them 
in the processes of string-brane scatterings.

For acquiring the on-shell scattering amplitude of
closed strings from a D-brane 
we should calculate overlap of an 
outgoing state and the boundary state, associated with 
the D-brane, via the closed string 
propagator and vertex operators.
Therefore, the tree-level scattering amplitude of $n + 2$ closed 
strings from a D-brane is given by \cite{5}-\cite{9},
\begin{eqnarray}
\mathcal{A}=
\langle V|c_{-1}\tilde{c}_{-1}
\int d^{2}z\int d^{2}z_{1}\mathcal{V}_{1}(z_{1},\bar{z}_{1})
\ldots \int d^{2} z_{n}\mathcal{V}_{n}(z_{n},\bar{z}_{n})
\mathcal{V}'(z,\bar{z})
(b_{0}+\tilde{b}_{0})\mathcal{D}|B\rangle,
\end{eqnarray}
where the integrals run over the upper 
half of the complex plane, and $\mathcal{D}$
is the closed string propagator. 
The ghost factor $b_{0}+\tilde{b}_{0}$ 
has been inserted to remove the
factor $(c_{0}+\tilde{c}_{0})/2$ of the 
ghost boundary state (2.7). 
The ghost modes $c_{-1}$ and ${\tilde c}_{-1}$
are also eliminated by the outgoing state and the state (2.7).
Hence, we don't need to worry about the ghost 
sector.

We can relocate the propagator to the left
and consider its effect on the outgoing state, hence, it 
disappears from the amplitude \cite{4}, \cite{5}.
Thus, we receive the following convenient amplitude
which is ghost-free 
\begin{eqnarray}
\mathcal{A}=\frac{\alpha'}{4 \pi}
\langle V_x|\int d^{2}z \int d^{2} z_{1}...
\int d^{2} z_{n} \mathcal{V}_{1}(z_{1},\bar{z}_{1})
...\mathcal{V}_{n}(z_{n},\bar{z}_{n})
\mathcal{V}'(z,\bar{z})
|B\rangle^{(0)}\otimes|B\rangle^{({\rm osc})}~,
\end{eqnarray} 
where $|V_x \rangle$ is the matter part of the outgoing 
state. In this formula the ranges of all 
integrals are outside of the 
unit circle, i.e. $|z_l|>1$ with 
$z_l \in \{z, z_1, \ldots, z_n\}$.

Note that 
from the point of view of the low-energy effective 
action a D-brane is a charged massive
object and, therefore, its presence  
inevitably induces a curvature into the spacetime.
However, the string-brane scattering amplitudes which are
calculated in the Minkowski spacetime comprise 
information for the dynamics  
in an effective curved spacetime, e.g. see \cite{2}.

\subsection{Elastic scattering of the Kalb-Ramond state}

Now we construct the amplitude concerning 
the elastic scattering of the Kalb-Ramond string state 
from a single dynamical-dressed D$p$-brane. 
Our calculations will be in the $t$-channel.
Using the characteristic feature
of the worldsheet duality
it is possible to recast the amplitude 
in the $s$-channel too. Note that 
when the vertex operators approach to each other
the $t$-channel case occurs. The $s$-channel appears 
when one of the vertex operators approaches to the
boundary of the string worldsheet.
 
The tree-level amplitude regarding the scattering 
of the Kalb-Ramond state from our D$p$-brane is specified by
\begin{eqnarray}
&~& \mathcal{A}_{\rm KR}=\frac{\alpha'}{4\pi}
\langle B_{\rm KR}|V_{\rm KR}^{(0,0)}(\zeta,k)
|B\rangle^{(0)}\otimes|B\rangle^{({\rm osc})}~.
\end{eqnarray}
The vertex operator, associated 
with this state in the (0,0) picture, has the structure 
\begin{eqnarray}
&~& V_{\rm KR}^{(0,0)}(\zeta,k)=
\int_{|z|>1}
d^{2}z \mathcal{V}_{\rm KR}^{(0,0)}(\zeta,k;z,{\bar z})~,
\nonumber\\
&~& \mathcal{V}_{\rm KR}^{(0,0)}(\zeta,k;z,{\bar z})
=\frac{\sqrt{2}\kappa}{\pi \alpha'}
\zeta_{\mu\nu}\partial X^{\mu} 
\bar{\partial} X^{\nu}e^{ik.X}~,
\end{eqnarray}
where the polarization tensor 
$\zeta_{\mu\nu}$ is antisymmetric, and 
$\kappa = (2\pi)^{7/2}(\alpha')^2 g_s/\sqrt{2}$ is
the gravitational constant and $g_s$ is the string coupling.
The amplitude (3.3) elaborates  
the physics of a single D$p$-brane which interacts with
two closed strings. 

By introducing the Kalb-Ramond state into
Eq. (3.3) the amplitude, for 
outgoing and incoming states with the momenta 
$k_1$ and $k_2$ and polarizations $\zeta_{(1)\mu\nu}$ and 
$\zeta_{(2)\mu\nu}$ respectively, possesses the form  
\begin{eqnarray}
\mathcal{A}_{\rm KR}= \frac{\alpha'}{4\pi}
\langle \textbf{1}_x|\left(V^{(0,0)}_{\rm KR}
(\zeta_{1},k_{1})\right)^{\dag} 
V^{(0,0)}_{\rm KR}(\zeta_{2},k_{2})
|B\rangle^{(0)}\otimes|B\rangle^{({\rm osc})}~,
\end{eqnarray}
where the vacuum $|\textbf{1}_x\rangle$ is the 
matter part of the total vacuum
$|\textbf{1}\rangle=
|\textbf{1}_x\rangle \otimes|\textbf{1}_{\rm gh}\rangle$.
Combining Eqs. (3.4) and (3.5) eventuates to the equation  
\begin{eqnarray}
\mathcal{A}_{\rm KR}&=&
\frac{\kappa^{2}}{2\pi^{3}\alpha'}
\zeta_{(1)\mu\nu}\zeta_{(2)\mu'\nu'}
\int d^{2}z_{1}\int d^{2}z_{2} 
\nonumber\\
&\times& \langle \textbf{1}_x|
\Big{[}e^{-ik_{1}\cdot X_{1}(z_{1},\bar{z}_{1})} 
(\bar{\partial_1} X_{L}^{\nu}(\bar{z}_{1}))^{\dag}
(\partial_1 X_{R}^{\mu}(z_{1}))^{\dag}\Big{]}
\nonumber\\
&\times& \Big{[} \partial_2 X_{R}^{\mu'}(z_{2})
\bar{\partial_2} X_{L}^{\nu'}(\bar{z}_{2})
e^{ik_{2}\cdot X_{2}(z_{2},\bar{z}_{2})}\Big{]}
|B\rangle^{(0)}\otimes|B\rangle^{({\rm osc})}.
\end{eqnarray}
This amplitude clearly is due to the two worldsheets 
that are semi-infinite cylinders.

From now on, for simplification, we consider 
a perpendicular incident and reflection 
of the incoming and outgoing states.
The equation of motion, in terms of the complex 
coordinates of the worldsheet, is
$\partial_z\partial_{\bar z}X^\mu(z ,{\bar z} )=0 $.
Using the closed string solution of it,
i.e. $X^\mu(z ,{\bar z} )=
X_{\rm R}^{\mu}(z)+X_{\rm L}^{\mu}(\bar{z})$, 
with  
\begin{eqnarray}
X_{\rm R}^{\mu}(z)&=&\frac{1}{2}x^{\mu}
-\frac{i}{2}\alpha'k^{\mu}\ln(z)
+i\sqrt{\frac{\alpha'}{2}}\sum_{m\neq 0}
\frac{\alpha^{\mu}_m}{m z^{m}}~,
\nonumber\\
X_{\rm L}^{\mu}(\bar{z})&=&\frac{1}{2}x^{\mu}
-\frac{i}{2}\alpha'k^{\mu}\ln(\bar{z})
+i\sqrt{\frac{\alpha'}{2}}\sum_{m\neq 0}
\frac{\tilde{\alpha}^{\mu}_m}{m \bar{z}^{m}}~,
\nonumber
\end{eqnarray}
and defining the operators 
\begin{eqnarray}
\eta_{r}&=&-\sqrt{\frac{\alpha'}{2}}
\sum_{m\neq 0}^{\infty}
\frac{k_{r}}{m}\cdot\frac{\alpha_{-m}}{\bar{z}_{r}^{m}}~,
\nonumber\\
\tilde{\eta}_{r}&=&-\sqrt{\frac{\alpha'}{2}}
\sum_{m\neq 0}^{\infty}
\frac{k_{r}}{m}\cdot
\frac{\tilde{\alpha}_{-m}}{z_{r}^{m}}~,
\end{eqnarray}
with $r=1,2$, Eq. (3.6) finds the form
\begin{eqnarray}
\mathcal{A}_{\rm KR}&=&\frac{\alpha'\kappa^{2}}{8\pi^{3}}
\;\zeta_{(1)\mu\nu}\zeta_{(2)\mu'\nu'}
\int d^{2}z_{1} 
\int d^{2}z_{2} \langle \textbf{1}_x|
\bigg{\{}e^{-ik_{1}\cdot x_{1}}
e^{\eta_{1}+\tilde{\eta}_{1}}
\nonumber\\
&\times& \bigg{[}-\sqrt{\frac{\alpha'}{2}}\sum_{m=1}^{\infty}
\left(\frac{k_{1}^{\nu}\alpha^{\mu}_{m}}{z_{1}\bar{z}_{1}^{-m+1}}+
\frac{k_{1}^{\nu}\alpha^{\mu}_{-m}}{z_{1}
\bar{z}_{1}^{m+1}}+
\frac{k_{1}^{\mu}\tilde{\alpha}^{\nu}_{m}}
{z_{1}^{-m+1}\bar{z}_{1}}+
\frac{k_{1}^{\mu}\tilde{\alpha}^{\nu}_{-m}}
{z_{1}^{m+1}\bar{z}_{1}}\right)
\nonumber\\
&-&\sum_{m_{1}=1}^{\infty}\sum_{m_{2}=1}^{\infty}
\left(\frac{\tilde{\alpha}^{\nu}_{m_1}
\alpha_{m_2}^{\mu}}{z_{1}^{-m_1+1}\bar{z}_{1}^{-m_2+1}}
+\frac{\tilde{\alpha}^{\nu}_{-m_1}
\alpha_{-m_2}^{\mu}}{z_{1}^{m_1+1}\bar{z}_{1}^{m_2+1}}
+\frac{\tilde{\alpha}^{\nu}_{m_1}
\alpha_{-m_2}^{\mu}}{z_{1}^{-m_1+1}\bar{z}_{1}^{m_2+1}}
+\frac{\tilde{\alpha}^{\nu}_{-m_1}
\alpha_{m_2}^{\mu}}{z_{1}^{m_1+1}\bar{z}_{1}^{-m_2+1}}
\right)\bigg{]}
\nonumber\\
&\times&\bigg{[}
-\sqrt{\frac{\alpha'}{2}}\sum_{m=1}^{\infty}
\bigg{(}\frac{k_{2}^{\nu'}\alpha^{\mu'}_{-m}}
{z_{2}^{-m+1}\bar{z}_{2}}+
\frac{k_{2}^{\nu'}\alpha^{\mu'}_{m}}{z_{2}^{m+1}\bar{z}_{2}}
+ \frac{k_{2}^{\mu'}\tilde{\alpha}^{\nu'}_{-m}}{z_{2}
\bar{z}_{2}^{-m+1}}+\frac{k_{2}^{\mu'}
\tilde{\alpha}^{\nu'}_{m}}{z_{2}\bar{z}_{2}^{m+1}}\bigg{)}
\nonumber\\
&-&\sum_{m_{1}=1}^{\infty}\sum_{m_{2}=1}^{\infty}
\left(\frac{\alpha_{-m_1}^{\mu'}
\tilde{\alpha}^{\nu'}_{-m_2}}{z_{2}^{-m_1+1}\bar{z}_{2}^{-m_2+1}}
+\frac{\alpha_{m_1}^{\mu'}
\tilde{\alpha}^{\nu'}_{m_2}}{z_{2}^{m_1+1}\bar{z}_{2}^{m_2+1}}
+\frac{\alpha_{m_1}^{\mu'}
\tilde{\alpha}^{\nu'}_{-m_2}}{z_{2}^{m_1+1}\bar{z}_{2}^{-m_2+1}}
+\frac{\alpha_{-m_1}^{\mu'}
\tilde{\alpha}^{\nu'}_{m_2}}{z_{2}^{-m_1+1}\bar{z}_{2}^{m_2+1}}
\right)\bigg{]}
\nonumber\\
&\times&e^{ik_{2}\cdot x_{2}}
e^{\eta^\dagger_{2}+\tilde{\eta}^\dagger_{2}}\bigg{\}}
|B\rangle^{(0)}\otimes|B\rangle^{({\rm osc})}~.
\end{eqnarray}

According to Ref. \cite{16} (Appendix 7.A)
and applying the oscillating part of the boundary 
state, we receive the identity 
\begin{eqnarray}
\langle \textbf{1}_x|e^{\eta_{1}
+\tilde{\eta}_{1}}e^{\eta_{2}^{\dag}
+\tilde{\eta}_{2}^{\dag}}|B'\rangle^{\rm (osc)}&=&
\exp \left[\langle \textbf{1}_x|\eta_{1}\tilde{\eta}_{1}
\eta_{2}^{\dag}\tilde{\eta}_{2}^{\dag}
|B'\rangle^{\rm (osc)}\right],
\end{eqnarray}
where the state $|B'\rangle^{\rm (osc)}$ is similar to
Eq. (2.4) without the overall infinite product.
The exponent part of the right-hand side 
is simplified as  
\begin{eqnarray}
&~& \langle \textbf{1}_x|\eta_{1}\tilde{\eta}_{1}
\eta_{2}^{\dag}\tilde{\eta}_{2}^{\dag}
\exp \left(-\sum_{n=1}^{\infty}
\frac{1}{n}\alpha_{-n}^{\mu}S_{(n)\mu\nu}
\tilde{\alpha}_{-n}^{\nu}\right)|\textbf{1}_x\rangle
\nonumber\\
&~& = -\frac{\alpha'^{2}}{4}
k_{1\mu}k_{1\nu}k_{2\mu'}k_{2\nu'}
\sum_{n=1}^{\infty}\left[
\frac{1}{n^{2}}
\left(S_{(n)}^{\mu\nu}S_{(n)}^{\mu'\nu'}
+S_{(n)}^{\mu\nu'}S_{(n)}^{\mu'\nu}\right)
\left(\frac{z_{1}\bar{z}_{1}}{z_{2}\bar{z}_{2}}\right)^{n}
\right].
\end{eqnarray}
Thus, the quantity in Eq. (3.9) takes the value  
\begin{eqnarray}
\exp \left[\langle \textbf{1}_x|\eta_{1}\tilde{\eta}_{1}
\eta_{2}^{\dag}\tilde{\eta}_{2}^{\dag}
|B'\rangle^{\rm (osc)}\right]=
\exp\left[-\frac{\alpha'^{2}}{4}
\sum_{n=1}^{\infty}
\frac{\lambda_{(n)}}{n^{2}}\left(\frac{z_{1}
\bar{z}_{1}}{z_{2}\bar{z}_{2}}\right)^{n}\right],
\end{eqnarray}
where 
\begin{eqnarray}
\lambda_{(n)}=k_{1\mu}k_{1\nu}k_{2\mu'}
k_{2\nu'}\left(S_{(n)}^{\mu\nu}S_{(n)}^{\mu'\nu'}
+S_{(n)}^{\mu\nu'}S_{(n)}^{\mu'\nu}\right).
\end{eqnarray}
In fact, the scattering of strings 
from the branes drastically provides 
reliable keys for extracting some essential quantities  
such as the sizes of the branes. Hence,
the exponential factor of  Eq. (3.11), which is a portion 
of the scattering amplitude, defines the characteristic length 
of the system as the order $\sqrt{\alpha'}$.
This length effectively indicates the thickness
of the target D-brane.

Eq. (3.11) implies that the scattering amplitude 
exponentially depends on the factors 
$\{\alpha'^2 \lambda_{(n)}/n^2\;| n\in \mathbb{Z}^+\}$. 
We shall compute the amplitude approximately,
i.e. we consider the limit
$\alpha' \rightarrow 0$ such that for 
all mode numbers of string the inequality  
$\alpha'^2 \lambda_{(n)}/n^2 << 1$ to be valid.

By applying a convenient choice for  
the positions of the vertex operators
as $z_{1}=iy$ and $z_{2}=i$, e.g. see Ref. \cite{37}
and \cite{38}, 
we can use the well-known integral representation 
of the Euler beta-function 
\begin{eqnarray}
\int_{0}^{1}t^{a-1}(1-t)^{b-1}dt
=\frac{\Gamma(a)\Gamma(b)}{\Gamma(a+b)}.
\end{eqnarray}
This enables us to simplify the integration
over the variable $y$.

We analyze the amplitude for two prominent cases:
the unstable target D-brane and the stable one.
These special cases are due to the presence and absence 
of the background tachyon field, respectively.
Note that the tachyon condensation phenomenon 
drastically imposes a collapse to the brane.

\subsubsection{Presence of the tachyonic background field}

In the presence of the tachyon field 
the amplitude (3.8) can be approximately written as
\begin{eqnarray}
\mathcal{A}_{\rm KR}& \approx &\mathcal{A}_{(0)\rm KR}
+\mathcal{A}_{(1)\rm KR}~,
\end{eqnarray}
where $\mathcal{A}_{(0)\rm KR}$ is the zero slope 
limit (i.e. $\alpha'\rightarrow 0$) part,
\begin{eqnarray}
\mathcal{A}_{(0)\rm KR}&=&\frac{\alpha'
\kappa^{2}T_p V_{p+1}\prod_{n=1}^{\infty}
\left[\det{\left(\mathcal{K}
-\mathcal{F}+\dfrac{i}{2n}U\right)}
\right]^{-1}}{2(2\pi)^{28-p}
\sqrt{\det (U/2)\det {\mathcal W}}} 
(\gamma_{\rm EM}-1)
\nonumber\\
&\times& 
\Big\{-\zeta_{(1)ij}\zeta_{(2)ij}
+ \sum_{n=1}^{\infty}\Big{[}
\zeta_{(1)i\alpha}\zeta_{(2)i\beta}
\left( S_{(n)}^{\alpha\beta}+ S_{(n)}^{\beta\alpha}\right)
\nonumber\\
&+&\zeta_{(1)\alpha\beta}
\zeta_{(2)\alpha'\beta'}\left(S_{(n)}^{\alpha\beta'}
S_{(n)}^{\alpha'\beta}
+S_{(n)}^{\alpha\beta}S_{(n)}^{\alpha'\beta'}\right)
\Big{]}\Big\},
\end{eqnarray}
where $\gamma_{\rm EM}=0.577 \cdots$ 
is the Euler-Mascheroni number which was
entered via a regularization scheme, and 
the matrix ${\mathcal W}_{\alpha\beta}$ has the definition  
\begin{eqnarray}
{\mathcal{W}}_{\alpha\beta} &=& 
\begin{cases} 
-i \alpha'(U^{-1}\mathcal{K}+ \mathcal{K}^T 
U^{-1})_{\alpha\beta},  
& \mbox{if }\alpha=\beta, \\
-2i \alpha'(U^{-1}\mathcal{K}+ \mathcal{K}^T 
U^{-1})_{\alpha\beta}, & \mbox{if }\alpha\neq\beta,
\end{cases} 
\nonumber\\
&=& i\alpha' \left[-2\left( U^{-1}\mathcal{K}
+\mathcal{K}^T U^{-1}\right)_{\alpha\beta}
+ \left( U^{-1}\mathcal{K}
+\mathcal{K}^T U^{-1}\right)_{\alpha\alpha}
\delta_{\alpha\beta}\right].
\end{eqnarray}
As we see in this approximation the 
amplitude completely is independent 
of the energies and momenta of the incoming
and outgoing strings.

The next $\alpha'$-correction of the 
amplitude has the feature 
\begin{eqnarray}
\mathcal{A}_{(1)\rm KR}&=&\frac{2
\kappa^{2}T_p V_{p+1}\prod_{n=1}^{\infty}
\left[\det{\left(\mathcal{K}
-\mathcal{F}+\dfrac{i}{2n}U\right)}
\right]^{-1}}{(2\pi)^{28-p}
\sqrt{\det (U/2)\det {\mathcal W}}}
\left(\frac{\alpha'}{2}\right)^{3/2}\gamma_{\rm EM}
\nonumber\\
&\times& \left( k^i_2 -k^i_1\right)
\Big\{-\zeta_{(1)ij}\zeta_{(2)jj'}k_{2}^{j'}
-\zeta_{(1) i j}\zeta_{(2)j0}k_{2}^{0}
\nonumber\\
&+& \Big{[} \zeta_{(1)\alpha\beta}
\zeta_{(2)i0}k_{2}^{0}+\zeta_{(1)i\beta}
\zeta_{(2)\alpha j}k_{2}^{j}
+\zeta_{(1)i\beta}\zeta_{(2)\alpha0}k_{2}^{0}
\Big{]}\left(\sum_{n=1}^{\infty} S_{(n)}^{\alpha\beta}\right)
\Big\}.
\end{eqnarray}
In fact, the string scattering gives rise a recoil to the brane,
which can be partly seen by the difference of the 
momenta in the amplitude, i.e. $k_{2}^{i}-k_{1}^{i}$.
Eqs. (3.15) and (3.17) elucidate that the tachyon 
field induces infinite partial amplitudes. This is
an effect of the scattering from an unstable brane.

We should note that derivations of Eqs. (3.15) and 
(3.17) generally are on the basis of the quantum mechanical 
techniques, specially we used the commutation relations
$[x^\mu , p^\nu]=i\eta^{\mu\nu}$ and 
$[\alpha^\mu_m , \alpha^\nu_n]=
[{\tilde \alpha}^\mu_m , {\tilde \alpha}^\nu_n]
=m\eta^{\mu\nu}\delta_{m+n,0}$. Besides,
some computational techniques of the Appendix 7.A of Ref. \cite{16}
have been applied. Since the scattering amplitude (3.14)
was approximately computed the explicit forms
of the $\Gamma$-functions disappeared.

\subsubsection{Absence of the tachyonic background field}

For a stable brane, i.e. 
in the absence of the tachyon field, 
the matrix $S_{(n)}^{\alpha\beta}$ and 
the infinite product in Eqs. (3.15) and 
(3.17) will be free of 
the string mode numbers. Therefore, 
the scattering amplitude possesses the form
\begin{eqnarray}
\mathcal{A'}_{\rm KR}& \approx &
\mathcal{A'}_{(0)\rm KR}
+\mathcal{A'}_{(1)\rm KR},
\end{eqnarray}
whit the ingredients 
\begin{eqnarray}
\mathcal{A'}_{(0)\rm KR}&=&\frac{\alpha'
\kappa^{2}T_p V_{p+1}\sqrt{\det
(\mathcal{K}-\mathcal{F})}}{2(2\pi)^{28-p}} 
(\gamma_{\rm EM}-1)
\nonumber\\
&\times& 
\Big\{-\zeta_{(1)ij}\zeta_{(2)ij}
+\zeta_{(1)i\alpha}\zeta_{(2)i\beta}
\left( S^{\alpha\beta}+ S^{\beta\alpha}\right)
\nonumber\\
&+&\zeta_{(1)\alpha\beta}
\zeta_{(2)\alpha'\beta'}\left(S^{\alpha\beta'}
S^{\alpha'\beta}
+S^{\alpha\beta}S^{\alpha'\beta'}\right)
\Big\},
\end{eqnarray}
\begin{eqnarray}
\mathcal{A'}_{(1)\rm KR}&=&\frac{2
\kappa^{2}T_p V_{p+1}\sqrt{\det
(\mathcal{K}-\mathcal{F})}}{(2\pi)^{28-p}} 
\left(\frac{\alpha'}{2}\right)^{3/2}\gamma_{\rm EM}
\nonumber\\
&\times& \left( k^i_2 -k^i_1\right)
\Big\{-\zeta_{(1)ij}\zeta_{(2)jj'}k_{2}^{j'}
-\zeta_{(1) i j}\zeta_{(2)j0}k_{2}^{0}
\nonumber\\
&+& \Big{[} \zeta_{(1)\alpha\beta}
\zeta_{(2)i0}k_{2}^{0}+\zeta_{(1)i\beta}
\zeta_{(2)\alpha j}k_{2}^{j}
+\zeta_{(1)i\beta}\zeta_{(2)\alpha0}k_{2}^{0}
\Big{]}S^{\alpha\beta}\Big\}.
\end{eqnarray}
The matrix $S^{\alpha\beta}$ has the definition 
\begin{eqnarray}
S^{\alpha\beta} = \left(\left(\mathcal{K}
-\mathcal{F}\right)^{-1}
\left(\mathcal{K}+\mathcal{F}\right)\right)^{\alpha\beta}.
\end{eqnarray}

As a special case let quench the magnetic part
of the total field strength, i.e.  
$\mathcal{F}_{{\bar \alpha}{\bar \beta}}=0$ with 
${\bar \alpha},{\bar \beta}\in \{1,2,\cdots,p\}$. 
Besides, stop the spatial rotation of the brane,
i.e. $\omega_{{\bar \alpha}{\bar \beta}}=0$.
Thus, the square root factor of the scattering 
amplitude reduces to $\sqrt{1-V^2-E^2
+2{\vec E}\cdot {\vec V}}$, where the 
components of the linear velocity 
of the brane and the total electric field are given by
$V_{\bar \alpha}=4\omega_{0{\bar \alpha}}$ and 
$E_{\bar \alpha}=\mathcal{F}_{0{\bar \alpha}}$.
We observe that for our setup the extra term 
$2{\vec E}\cdot {\vec V}$ is nonzero, while  
it is absent for 
the conventional transverse dynamics of the branes.
However, by adjusting the electric field and 
brane velocity such that $|{\vec E}- {\vec V}|\rightarrow 1$
the scattering amplitude obviously goes to zero. Similarly,
for the case ${\vec E} - {\vec V}\rightarrow {\vec 0}$ 
the prefactor of the amplitude tends to its maximum value.

Scatterings from both unstable and 
stable branes manifestly demonstrate 
that the polarization elements $\zeta_{(1)ij}$ and 
$\zeta_{(2)ij}$ do not mix
with the matrices $S^{\alpha \beta}_{(n)}$ and 
$S^{\alpha \beta}$. This implies that if the 
incident state has a polarization with only 
$\zeta_{(1)ij} \neq 0$ then the matrices 
$S^{\alpha \beta}_{(n)}$ and $S^{\alpha \beta}$ 
do not appear in the scattering amplitudes. Physically 
this means that an incoming state 
with this special polarization 
cannot completely explore the 
structure of the target brane.

\section{The DBI-like actions}

The Dirac-Born-Infeld (DBI) action and its 
extended versions  
are the low energy effective actions
of the tachyon and massless fields. In other words, these
actions elaborate the interactions 
between the foregoing fields and the corresponding D-brane.
We observe that the square root factor in Eqs. (3.19) 
and (3.20) clearly indicates a generalized 
DBI Lagrangian, associated with the stable D-brane.
The corresponding DBI-like action, by including the 
dilaton field, possesses the feature 
\begin{eqnarray}
S^{(\omega)}_{\rm DBI}&=& -T_p\int d^{p+1} \xi e^{-\phi}
\sqrt{-\det \left[{\tilde G}_{\alpha\beta}
-{\tilde B}_{\alpha\beta}+F_{\alpha\beta}
+4\omega_{\alpha\beta}\right]}~,
\end{eqnarray}
where ${\tilde G}_{\alpha\beta}$ and
${\tilde B}_{\alpha\beta}$ are pullbacks
of $G_{\mu\nu}$ and $B_{\mu\nu}$ on the
brane worldvolume, respectively.
This action explicitly comprises the effect of 
the brane dynamics. In the static gauge 
$\{\xi^\alpha=x^\alpha|\alpha=0,1,\cdots,p\}$
we obtain ${\tilde G}_{\alpha\beta}=\eta_{\alpha\beta}$
and ${\tilde B}_{\alpha\beta}=B_{\alpha\beta}$.
Hence, in this gauge for $\phi=0$ and 
constant $B_{\alpha\beta}$
and $F_{\alpha\beta}$ the Lagrangian in Eq. (4.1),
up to a constant factor, reduces to the 
square root factor of Eqs. (3.19) and (3.20).

In the same way, the prefactor of Eqs. (3.15) and (3.17) 
indicates the following effective action 
\begin{eqnarray}
S^{(\omega,T)}_{\rm DBI}&=& -T_p\int d^{p+1} \xi e^{-\phi}V(T)
\sqrt{-\det \left[{\tilde G}_{\alpha\beta}
-{\tilde B}_{\alpha\beta}+F_{\alpha\beta}
+4\omega_{\alpha\beta}\right]}\;
\mathcal{G}( {\bar U})~,
\end{eqnarray}
where our tachyon profile implies ${\bar U}_{\alpha\beta}
=2\left(\partial_\alpha T\partial_\beta T+
T\partial_\alpha \partial_\beta T\right)$.
The functional $\mathcal{G}( {\bar U})$ is given by 
\begin{eqnarray}
\mathcal{G}( {\bar U})=\frac{i(-2i\alpha')^{(p+1)/2}}
{\sqrt{\det ({\bar U} {\bar {\mathcal W}})}}
\det \Gamma\left[{\bf 1}+\frac{i}{2} 
\left({\tilde G}-{\tilde B}+F+4\omega
\right)^{-1}{\bar U}\right].
\end{eqnarray}
The symbol ``$\Gamma$'' represents the gamma-function, and 
the matrix ${\bar {\mathcal W}}$ possesses the form (3.16)
with ${\bar U}$ instead of $U$. The variable $V(T)$
is the tachyon potential, and the literature has  
proposed several forms for it. 
For a non-BPS brane it is well-known that the tachyon potential
is an even functional of the tachyon field $T$, and 
as $T \rightarrow \pm\infty$ it vanishes. Note that 
for acquiring this action we applied the following 
regularization schemes 
\begin{eqnarray}
&~& \prod^\infty_{n=1}a \rightarrow \frac{1}{\sqrt{a}},
\nonumber\\
&~& \prod^\infty_{n=1}\left(1+\frac{a}{n}\right)
\rightarrow \frac{1}{\Gamma (a+1)}.
\end{eqnarray}

We observe that the kinetic term of the tachyon field 
is not under the square root. There are various 
effective actions in which the kinetic term of 
the tachyon is out of the square root,
e.g. see \cite{39, 40, 41, 42}.
In addition to these features, there are some 
other extended forms for the 
tachyonic part of the actions. For example, 
some of the generalized actions can be found 
in the Refs. \cite{39, 40, 41, 42, 43, 44, 45, 46, 47}. 
However, for a stationary brane and 
in the absence of the massless fields the action (4.2)
reduces to an effective action for the tachyon field.
Similarly, in the absence of the tachyon field and
for the tachyon potential with $V(0)=1$ the action 
(4.2) accurately reduces to the action (4.1), as expected.

As we know the physical tension of 
a D-brane is proportional to the  
inverse of the closed string coupling $g_s$. 
This fact implies that the D-branes are essential part of the 
non-perturbative string theory.
Therefore, the brane effective actions (4.1) and (4.2), 
which originated 
from the string-brane scattering, clarify that such
scatterings prepare a remarkable
intuition on the non-perturbative string theory.

Note that the unstable branes
under the tachyon condensation phenomenon
collapse into the closed string vacuum or drastically 
decay to the stable configurations with lower dimensions
\cite{34, 47, 48, 49}. 

\section{Conclusions}

For calculating the scattering amplitude 
we applied the string operator formalism.
In this reliable method an amplitude is computed by 
evaluating the correlation function 
of the vertex operators which are corresponding
to the string states,
presented in the scattering process.

We acquired the scattering amplitude 
of the Kalb-Ramond state from a D$p$-brane with the 
following background fields: a $U(1)$ gauge 
potential with constant field strength,  
a constant antisymmetric field and a
quadratic tachyon field. The brane 
has a rotation and a linear motion. 
The amplitude extremely 
depends on the background fields and the brane dynamics.
The variety of the input parameters 
$\{B_{\alpha\beta}, F_{\alpha\beta}, U_{\alpha\beta};
\omega_{\alpha\beta};p\}$
gave a general feature to the amplitude. The strength of 
the scattering can be accurately adjusted by
these variables. For example, for a stable target brane
with an internal electric field and a linear 
motion this strength was
adjusted to the zero value and to the maximum value.

The scattering from an unstable 
brane was represented by infinite partial amplitudes.
Scatterings from both kind of the stable 
and unstable branes elucidate 
that for a particular polarization 
of the incoming state the scattering amplitudes 
are independent of the worldvolume matrices 
$S^{\alpha \beta}_{(n)}$ and $S^{\alpha \beta}$, 
which include information about the target branes.
In other words, structure of a target D-brane
can be partially investigated by such incident states.

We observed that at the zero slope limit
the scattering amplitude 
is independent of the energies and momenta
of the incoming and outgoing string states.   
By adding the $\alpha'$-corrections
dependence on these quantities is restored.
Besides, we received the characteristic
length of the system as the order $\sqrt{\alpha'}$,
which can be interpreted as the effective thickness
of the target brane.

As the final result, the scattering 
amplitudes enabled us to receive 
effective actions which are corresponding to the stable 
and unstable dynamical target D$p$-branes.
For the unstable brane, due to the initial 
tachyon profile, the kinetic term of the tachyon 
field was recast in a complicated functional.
These actions, because of the essential 
role of the D-branes, may shed light on the non-perturbative
string theory, and may possibly lead to 
a deeper understanding of the D-branes substantial
properties. 



\begin{thebibliography}{99}

\bibitem{1}
G. D'Amico, R. Gobbetti, M. Kleban and M. Schillo, 
JHEP \textbf{01}(2015) 050.

\bibitem{2}
G. D'Appollonio, P. Di Vecchia, R. Russo and G. Veneziano,
JHEP \textbf{1011} (2010) 100;
JHEP \textbf{2013} (2013) 126;
D. Amati, M. Ciafaloni and G. Veneziano, Phys. Lett.
\textbf{B 197} (1987) 81.

\bibitem{3}
R. Klebanov and L. Thorlacius, 
Phys. Lett. \textbf{B 371} (1996) 51-56.

\bibitem{4}
M. Frau, I. Pesando, S. Sciuto, A. Lerda and R. Russo,
Phys. Lett. \textbf{B 400} (1997) 52.

\bibitem{5}
M. Billo, P. Di Vecchia, M. Frau, A. Lerda, 
I. Pesando, R. Russo and S. Sciuto, Nucl. Phys. 
{\bf B 526} (1998) 199;
B. Craps, "D-branes and boundary states
in closed string theories" Ph.D. thesis; 
arXiv:hep-th/0004198.

\bibitem{6}
M. Bershadsky and D. Kutasov, Nucl. Phys. 
\textbf{B 382} (1992) 213-228.

\bibitem{7}
M. R. Garousi and R. C. Myers, 
Nucl. Phys. \textbf{B 475} (1996) 193-224.

\bibitem{8}
A. Hashimoto and I.R. Klebanov, 
 Phys. Lett. \textbf{B 381} (1996) 437-445.

\bibitem{9}
A. Hashimoto and I. Klebanov, Nucl.
Phys. Proc. Suppl. \textbf{55B} (1997) 118.

\bibitem{10}
Sh. Hirano and Y. Kazama,  Nucl.Phys. 
\textbf{B 499} (1997) 495-515.

\bibitem{11}
Y. Hikida, H. Takayanagi and T. Takayanagi, 
JHEP \textbf{0304} (2003) 032.

\bibitem{12}
M. Bianchi and P. Teresi, 
JHEP \textbf{1201} (2012) 161.

\bibitem{13}
D. Tong, JHEP \textbf{0602} (2006) 030.

\bibitem{14}
J. Maharana, Nuc. Phys. \textbf{B 896} ( 2015) 657-681.

\bibitem{15}
S. Stieberger, T. R. Taylor, 
Nuc. Phys. \textbf{B 903} (2016) 104-117.

\bibitem{16}
M. Green, J. Schwarz and E. Witten, 
``\textit{Superstring Theory}'', Vol. I,
Cambridge University Press, (1987).

\bibitem{17} 
C. G. Callan, I. R. Klebanov, 
Nucl. Phys. \textbf{B 465} (1996) 473.

\bibitem{18} 
M.B. Green and P. Wai, Nucl. Phys. 
\textbf{B431} (1994) 131.

\bibitem{19}
C. Bachas, Phys. Lett. \textbf{B 374} (1996) 37.

\bibitem{20}
M. Li, Nucl. Phys. \textbf{B 460} (1996) 351.

\bibitem{21} 
M.B. Green and M. Gutperle, Nucl. 
Phys. \textbf{B476} (1996) 484.

\bibitem{22} 
M. Frau, A. Liccardo and R. Musto, Nucl. Phys. 
\textbf{B 602} (2001) 39.

\bibitem{23} 
M. Billo, D. Cangemi, P. Di Vecchia, 
Phys. Lett. \textbf{B 400} (1997) 63.

\bibitem{24}
F. Hussain, R. Iengo and C. Nunez, 
Nucl. Phys. \textbf{B 497} (1997) 205.

\bibitem{25}
P. Di Vecchia, M. Frau, I. Pesando, 
S. Sciuto, A. Lerda and R. Russo, Nucl. 
Phys. \textbf{B507} (1997) 259.

\bibitem{26} 
C. G. Callan, C. Lovelace, C. R. Nappi, S. A. Yost,
Nucl. Phys. \textbf{B 288} (1987) 525;  
Nucl. Phys. \textbf{B 308} (1988) 221.

\bibitem{27} 
O. Bergman, M. Gaberdiel and 
G. Lifschytz, Nucl. Phys. \textbf{B509} (1998) 194.

\bibitem{28} 
T. Kitao, N. Ohta, J. G. Zhou, 
Phys. Lett. \textbf{B 428} (1998) 68.

\bibitem{29} 
S. Gukov, I. R. Klebanov, A. M. Polyakov, 
Phys. Lett. \textbf{B 423} (1998) 64.

\bibitem{30}
M. Bertolini, P. Di Vecchia, M. Frau, A. Lerda and R. 
Marotta, Nucl. Phys. \textbf{B 621}
(2002) 157.

\bibitem{31}
P. Di Vecchia, A. Liccardo, R. Marotta and 
F. Pezzella, Int. J. Mod. Phys. \textbf{A 20}
(2005) 4699-4796.

\bibitem{32}
H. Arfaei and D. Kamani, Phys. Lett. \textbf{B 452} (1999) 54,
hep-th/9909167;
Nucl. Phys. \textbf{B 561} (1999) 57-76, hep-th/9911146;
Phys. Lett. \textbf{B 475} (2000) 39-45, hep-th/9909079;
D. Kamani, Phys. Lett. \textbf{B 487} (2000) 187-191,
hep-th/0010019;
Annals of Physics \textbf{354} (2015) 394-400,
arXiv:1501.02453[hep-th];
Nucl. Phys. {\bf B 601} (2001) 149-168, hep-th/0104089;
Mod. Phys. Lett. \textbf{A 17} (2002) 237, hep-th/0107184;
Mod. Phys. Lett. \textbf{A 15} (2000) 1655, hep-th/9910043;
F. Safarzadeh-Maleki and D. Kamani,
Phys. Rev. \textbf{D 90} (2014) 107902, arXiv:1410.4948[hep-th];
Phys. Rev. \textbf{D 89} (2014) 026006, arXiv:1312.5489[hep-th];
M. Saidy-Sarjoubi and D. Kamani, Phys. Rev. \textbf{D 92} (2015)
046003, arXiv:1508.02084[hep-th].

\bibitem{33}
E. Maghsoodi and D. Kamani, 
Nucl. Phys. \textbf{B 922} (2017) 280-292, arXiv:1707.08383[hep-th].

\bibitem{34}
D. Kutasov, M. Marino and G. Moore, JHEP {\bf 0010} (2000) 045.

\bibitem{35}
E. T. Akhmedov, M. Laidlaw and G. W. Semenoff, JETP Lett. {\bf 77} 
(2003) 1-6; M. Laidlaw and G. W. Semenoff, 
JHEP {\bf 0311} (2003) 021.

\bibitem{36}
C.G. Callan, C. Lovelace, C.R. Nappi, S.A. Yost, Nucl.
Phys. {\bf B 293} (1987) 83.

\bibitem{37}
D. Friedan, E. Martinec and S. Shenker,
Nucl. Phys. {\bf B271} (1986) 93.

\bibitem{38}
D. Lust and S. Theisen, ``{\it Lectures on String
Theory}'', Springer-Verlag (1989).

\bibitem{39}
A. Sen, Phys. Rev. {\bf D 68} (2003) 066008.

\bibitem{40}
S. Sugimoto and S. Terashima, JHEP {\bf 0207} (2002) 025.

\bibitem{41}
N.D. Lambert and I. Sachs, Phys. Rev. {\bf D 67} (2003) 026005;
JHEP {\bf 0106} (2001) 060.

\bibitem{42}
S. Terashima and T. Uesugi, JHEP {\bf 0105} (2001) 054. 

\bibitem{43}
J.A. Minahan and B. Zwiebach, JHEP {\bf 0102} (2001) 034.

\bibitem{44}
J. A. Minahan, JHEP {\bf 0207} (2002) 030.

\bibitem{45}
D. Kutasov, M. Marino and G.W. Moore, 
``{\it Remarks on tachyon condensation in superstring 
field theory}'', arXiv:hep-th/0010108.

\bibitem{46}
N.D. Lambert and I. Sachs, JHEP {\bf 0106} (2001) 060.

\bibitem{47}
P. Kraus and F. Larsen, Phys. Rev. {\bf D 63} (2001) 106004.

\bibitem{48}
A. Sen, Int. J. Mod. Phys. {\bf A 14} (1999) 4061; 
JHEP {\bf 9808} (1998) 010. 

\bibitem{49}
E. Witten, Phys. Rev. {\bf D 47} (1993) 3405;
Phys. Rev. {\bf D 46} (1992) 5467.

\end{thebibliography}
\end{document}